\documentclass[runningheads,citeauthoryear]{apinv}
\usepackage{epsfig,cite,graphics}

\usepackage[T2A]{fontenc}
\usepackage[cp1251]{inputenc}
\usepackage[bulgarian,english]{babel}

\def\Ha {H$\alpha$~}

\begin{document}

\title{Wind structure in late-B supergiants}
\titlerunning{Wind structure in late-B SGs}
\author{Nevena Markova\inst{1} \and Haralambi Markov\inst{1}}
\authorrunning{Markova and Markov}
\tocauthor{Nevena Markova}
\institute{Institute of Astronomy, NAO, BAS 
	\newline
	\email{nmarkova@astro.bas.bg}
}
\papertype{contribution}
\maketitle
\begin{abstract}
Extended spectroscopic datasets of several late-B stars 
of luminosity class Ia revealed the presence of similar  
peculiarities in their \Ha profiles, which might be 
interpreted as indications of deviation from spherically 
symmetric, smooth wind approximation. Surface structures due 
to non-radial pulsations or weak, large-scale, dipole magnetic 
fields might be responsible for creating wind structure in the 
envelopes of these stars. 
\end{abstract}


\section*{Introduction}

The key limiting assumptions incorporated within current hot 
star model atmospheres include a globally stationary and 
spherically symmetric stellar wind with a smooth density 
stratification. Although these models are generally quite 
successful in describing the overall wind properties, there
are numerous observational and theoretical studies, which 
indicate that hot star winds are certainly not smooth and 
stationary. Most of the time-dependent constraints 
refer however to O-stars and early B supergiants (SGs), while 
mid- and late-B candidates are currently under-represented in 
the sample of stars investigated to date. 

Indeed, theoretical predictions supported by observational 
results (Markova and Puls \cite{MP}) indicate that while winds 
in late-B SGs are significantly weaker than those in O SGs, 
there is no currently established reason to believe that weaker 
winds might be less structured than stronger ones. 

\section{Results and discussion}

\begin{figure}[t]
\begin{minipage}{6.8cm}
\resizebox{\hsize}{!}
{\includegraphics{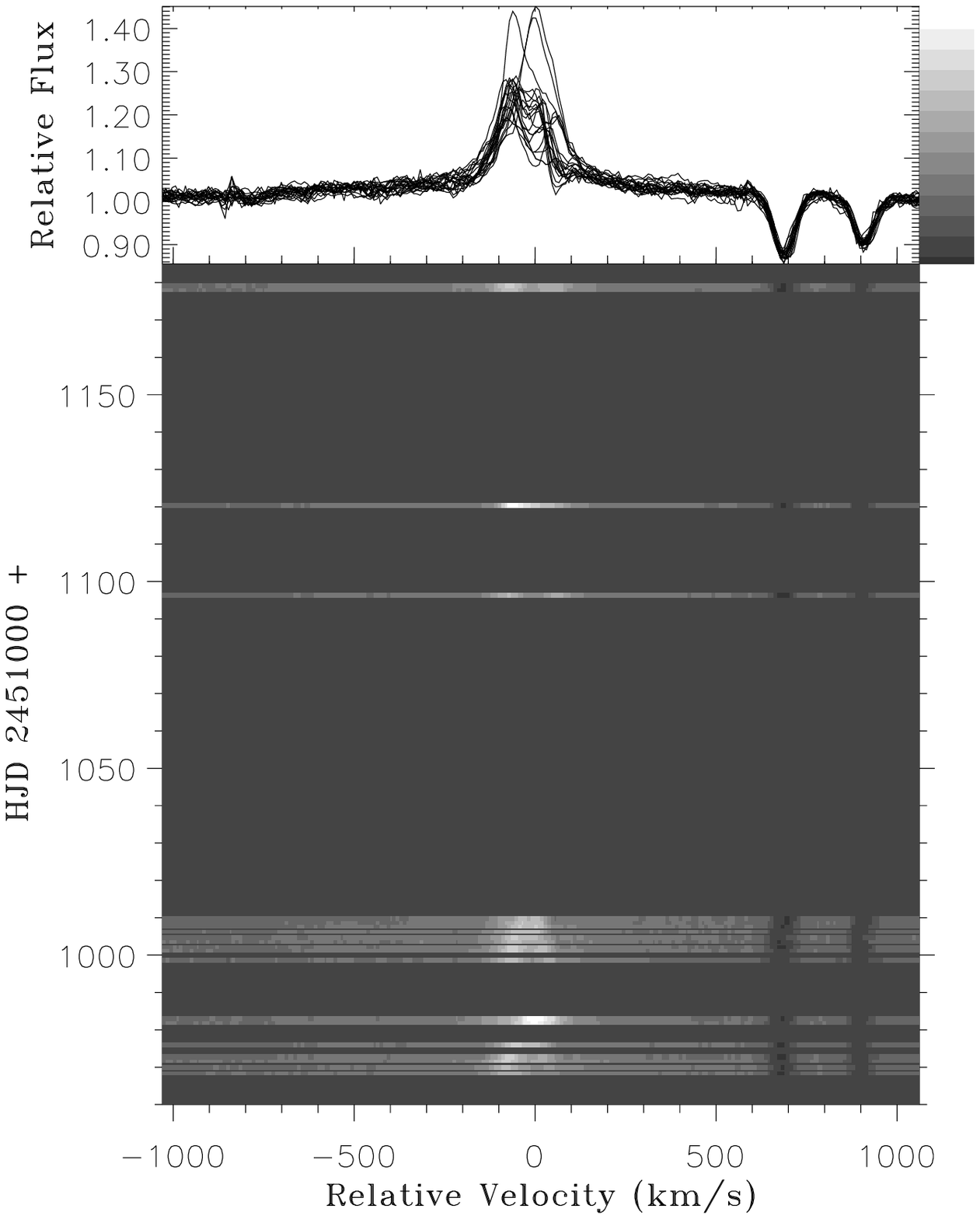}}
\end{minipage}
\hfill
\begin{minipage}{6.8cm}
\resizebox{\hsize}{!}
{\includegraphics{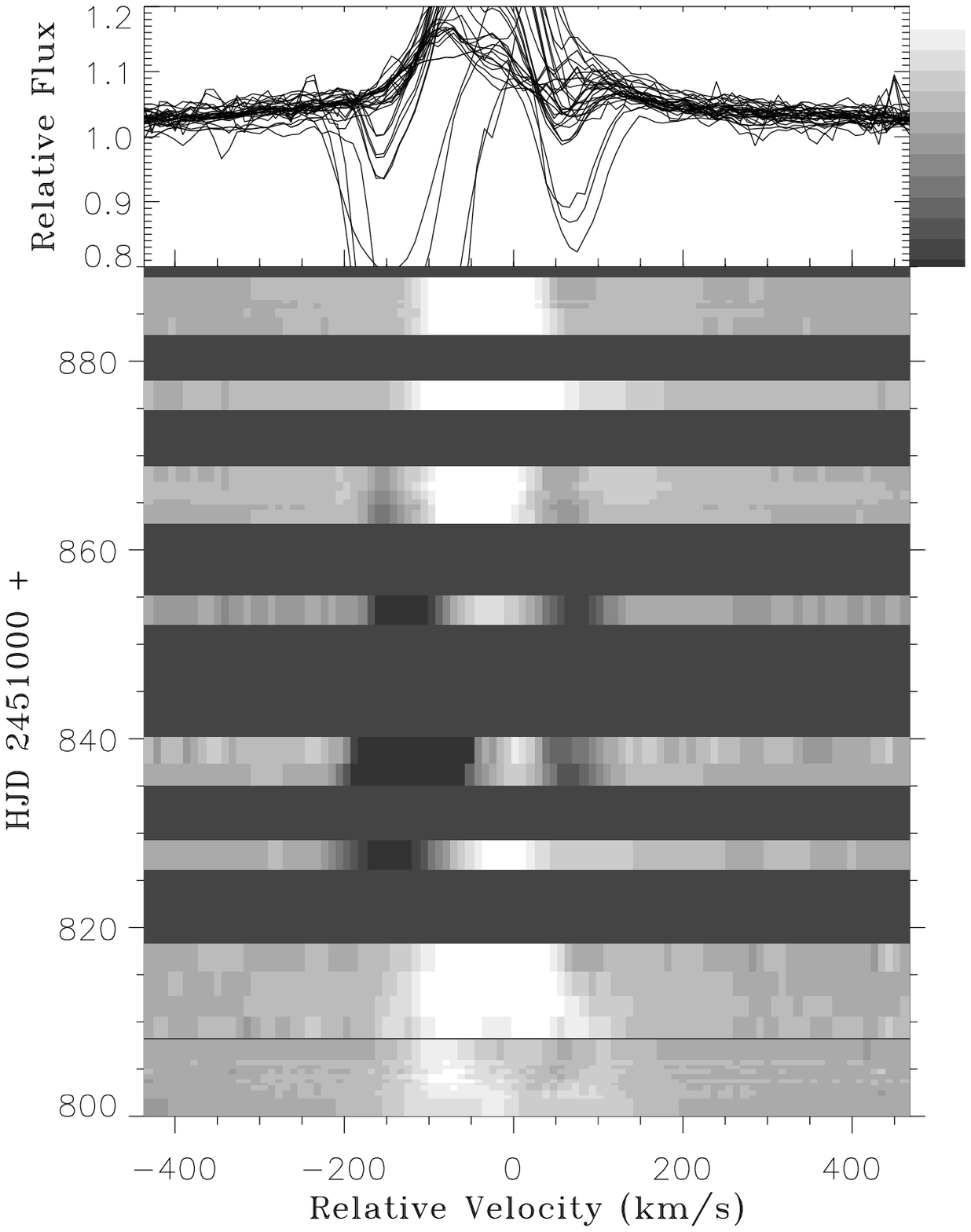}}
\end{minipage}
    \caption[]{Examples for typical variability in \Ha of HD~199\,478 
(B8 Iae) (from Markova et al. \cite{markova08})}
\label{hd199478}
\end{figure}

Long-term monitoring campaign of several late-B SGs, 
namely HD~199\,478 (Markova and Valchev \cite{MV}, 
Markova et al. \cite{markova08}), HD~91\,619, 
HD~43\,085 and  HD~96\,919 (Kaufer et al. 
\cite{kaufer96a, kaufer96b, Kaufer97}, Israelian et al. \cite
{Israel97}) revealed the presence of photometric and 
wind variability of quite similar signatures in their spectra.
In particular, the wind variability, as traced by H$\alpha$, is 
characterised by extremely strong, double-peaked emission 
with  V/R variations and occasional episodes of strong absorption 
with blue- and red-shifted features indicating simultaneous 
mass infall and outflow. (A typical example of such behaviour 
is given in Figure~\ref{hd199478}). 

Such line signatures cannot be reproduced in terms of the 
conventional (i.e. non-rotating, spherically symmetric, smooth) 
wind models, which instead predict profiles in absorption partly  
filled in by emission for SGs at this temperature regime 
(Markova et al. \cite{markova08}). Subsequently, axially symmetric, 
disc-like envelopes (Kaufer et al. \cite{kaufer96a}, Markova and 
Valchev \cite{MV}) and episodic, azimuthally 
extended, density enhancements in the form of co-rotating spirals 
rooted in the photosphere (Kaufer et al. \cite{kaufer96b}) or 
closed magnetic loops similar to those in our Sun (Israelian et al. 
\cite{Israel97}) have been assumed to account for the peculiar 
behaviour of \Ha in these stars.

In general, there are at least three possible ways to break 
the spherically symmetric wind geometry  and create large-scale 
winds structure around hot stars: by fast rotation, by surface 
structures and by large-scale, dipole magnetic fields.

\paragraph{Wind structure due to fast rotation}
Model calculations from the early 1990s (e.g. Bjorkman and 
Cassinelli \cite{BC93}) showed that $if$ the rotational rate 
of a hot star is above a given threshold determined by 
the ratio of its terminal wind velocity to the escape 
velocity, stellar rotation might converge the radiative 
driven wind flow towards the equator, creating a dense 
equatorial disc. However, observations indicate that even 
in fast rotating Be stars, this requirement is not 
fulfilled. In addition, our stars are not fast rotators: 
their rotational speeds are a factor of 3 to 5 lower than 
the corresponding  critical values. Thus, the fast rotation 
hypothesis can be rejected as a possible cause for wind 
structures in late-B SGs. 

\begin{figure}[t]
\centering
{\epsfig{file=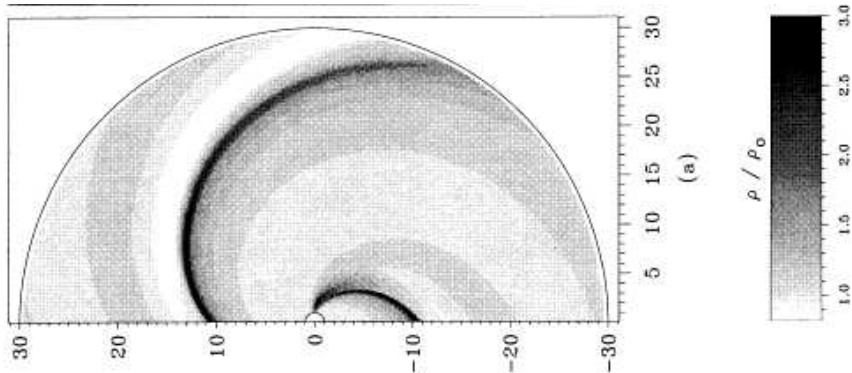, width=0.9\textwidth}}
    \caption[]{CIR structure created by a ``bright" spot on 
the surface of a rotating O star. (from Cranmer and Owocki 
\cite{CO96})}
\label{CIR}
\end{figure}

\paragraph{Surface structures}
Non-radial pulsations (NRPs) and magnetic fields might equally 
be responsible for driving the stellar surface into regions of 
different properties (Fullerton et al. \cite{Full96}). 
Results of 2D hydrodinamical simulations (Cranmer and Owocki 
\cite{CO96}) showed that
``bright/dark" spots on the stellar surface can effectively 
enhance/reduce the radiative driving, leading to the formation
of high/low-density, low/high-speed streams. Consequently, a 
specific wind structure, called Corotating Interaction Region 
(CIR) structure, forms where fast material collides with  
slow material giving rise to travelling features in various 
line diagnostics (e.g. Discrete Absorption Components in UV 
resonance lines of O stars, see e.g. Kaper et al. \cite{kaper96}). 
The CIR scenario for the case of a ``bright" surface spot 
in a rotating O star is schematically illustrated in 
Figure~\ref{CIR}.

Concerning the four late-B SGs considering here, non-radial 
pulsations due to $g$-modes oscillations have been suggested 
to explain absorption $lpv$ in their spectra (Kaufer et al. 
\cite{Kaufer97}, Markova and Valchev \cite{MV}, Markova et al. 
\cite{markova08}). This possibility is partially  supported 
by results from recent quantitative spectral analyses, which 
indicate that on the HR diagram, and for parameters derived 
with FASTWIND (Puls et al. \cite{P05}), 
these stars fall exactly in the region occupied 
by known variable B SGs, for which $g$-modes instability was 
suggested (Markova et al. \cite{markova08}). Also, the 
photometric behaviour of some of our targets  
(e.g. HD~199\,478, Percy et al. \cite{PAM}) seems 
to be consistent with a possible origin in terms of $g$-mode 
oscillations. Thus, it seems very likely that these stars are 
non-radial pulsators and therefore, may create, at least 
theoretically, wind structures via the CIR scenario described 
above. This possibility however has to be observationally proven.
In this respect, we note that no clear evidence of any causality 
between photospheric and wind (as traced by \Ha) variability has 
been derived so far for any of our targets (Kaufer et al. 
\cite{Kaufer97}, Markova et al. \cite{markova08}). Also, the 
variability patterns observed in their 
\Ha profiles do not give any evidence of migrating red-to-blue 
features, as those expected to originate from a CIR structure.

\paragraph{Dipole magnetic fields}
The possibility that magnetic fields can be responsible for 
the appearance of large-scale wind structures  in hot stars 
has been supported by recent magneto-hydrodynamical 
(MHD) simulations. Early results derived via such simulations 
(Babel and Montmerle \cite{Babel97}, Donati et al. 
\cite{Donati01}) indicated that a co-rotating, 
equatorial disc can be created around {\it non-rotating}, hot, 
main sequence stars due to a relatively weak bipolar magnetic 
field (about several KGauss). In this model, called  Magnetically 
Confined Wind Shock (MCWS) model, supersonic wind-streams from 
the two hemispheres are magnetically confined and directed 
towards the magnetic equatorial plane, where they collide and 
produce a strong shock giving rise to  X-ray emission.
\begin{figure}[t]
\centering
{\epsfig{file=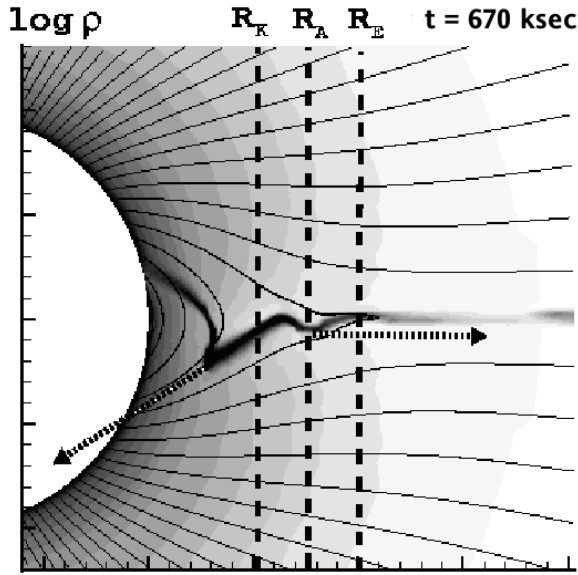, width=0.6\textwidth}}
    \caption[]{Density stratification for a model  with
stellar and wind parameters typical for O stars 670 ksec after 
the initial introduction of a dipole magnetic field. The arrows 
illustrate the upward and downward flow direction of dense 
material above and below the Keplerian radius (from ud-Doula and 
Owocki \cite{OD03}). The results for late-B SGs models are 
qualitatively similar.}
\label{magn_field}
\end{figure}

The MCWS model has been questioned by more recent simulations  
(ud-Doula and Owocki \cite{DO02}) which showed that  without 
any rotational support 
the material trapped within the magnetic loops would simply fall 
back along the field line to the loop foot-point, i.e. an infall of 
material in the form of dense knots, rather than an equatorial 
disc, would be generated. Additional MHD simulations  for $rotating$ 
hot stars with a magnetic dipole aligned to the stellar 
rotation axis furthermore indicated that depending on the 
magnetic spin-up an equatorial compression dominated 
by radial infall and/or outflows, with no apparent tendency to 
form a steady, Keplerian disc, might be created (Owocki and ud-Doula \cite{OD03}, ud-Doula et al. \cite{DOT08}). 

Due to their radiative envelopes normal (i.e. without any chemical peculiarities) B stars  are expected to be non-magnetic objects. 
Nonetheless,  during the last decade a growing number of direct 
observational evidence has been derived which indicates that 
relatively strong, stable, large-scale dipole magnetic fields  
do present in some B stars (e.g. SPB, Be,  $\beta$ Cep) 
(Henrichs et al. \cite{Henrichs00}, Neiner et al. \cite{Neiner01}, 
Bychkov et al. \cite{Bychkov}, Hurbig et al. \cite{Hurbig05, 
Hurbig07}). 

From the above outlined it appears that in, at least some, 
hot stars magnetic fields can be an alternative source of 
wind perturbations and asymmetries. And although the 
four late-B SGs discussed here have not been recognised 
so far as magnetically active stars (except for HD~34\,085, 
where a magnetic field of about 130$\pm$20~G was detected  
by Severny \cite{severny}), the potential role of magnetic 
fields in these stars remains intriguing, especially because 
it might provide a clue to 
understand the puzzling problem of the simultaneous presence 
of red- and blue-shifted absorptions/emissions in their 
\Ha profiles. 

To test this possibility new MHD simulations for the case of 
mid/late B SGs have been recently initiated. 
The preliminary results  (private communication, Asif ud-Doula) 
indicate that a pure dipole magnetic field of only a few 
tens of Gauss is required to obtain a $cool$ equatorial 
compression (with mass infall and outflow) around a 
rotating star with stellar and wind properties as 
derived with FASTWIND for HD~199\,478 (Markova and 
Puls \cite{MP}). Interestingly, few hundreds {\em ksec} 
after the onset of the magnetic field, the obtained 
density stratification in this late-B SGs model 
turned out to be qualitatively similar to that 
obtained for models with stellar and wind parameters typical
for O stars and early B SGs (see Figure~\ref{magn_field}). 

An obvious advantage of the model described above is that 
it allows  to interprete, at least qualitatively, 
some of the peculiar characteristics of \Ha in our targets. 
In particular, the presence of red/blue-shifted absorptions 
might be explained if one assumes that, due to some reasons, 
the plasma in the infalling or outflowing zones of the 
compression or in both of them (during the High Velocity 
Absorption episodes) 
can become optically thick  in the $Lyman$ continuum and 
L$_{\alpha}$. Then, \Ha will start to behave as a resonance 
line, i.e. to absorb and emit line photons (for more details 
see Markova et al. \cite{markova08}). The kinematic 
properties of the resulting absorption features is difficult 
to predict from simple qualitative considerations 
but it is in advance clear that these properties cannot be be 
dominated by stellar rotation (Townsend and Owocki \cite{TO05}). 

Concerning the interpretation of the peculiar \Ha emission, 
the situation is more complicated since such emission can 
originate from different parts of the envelope, under quite 
different physical conditions. More detailed 
quantitative analysis is required to check  all 
possibilities and investigate them further.

{\it Acknowledgements:}
This work was in part supported by the National Scientific 
Foundation to the Bulgarian Ministry of Education and Science 
(F-1407/2004).


\end{document}